\documentclass[twocolumn,english,prl,floatfix,citeautoscript,nofootinbib,superscriptaddress]{revtex4}
\usepackage{amsbsy}
\usepackage{latexsym,epsfig,graphicx}
\usepackage{dcolumn}
\usepackage{subfigure}
\usepackage{comment}
\usepackage{color}
\usepackage[colorlinks,urlcolor=blue,citecolor=blue]{hyperref}
\usepackage{amstext}
\usepackage{amssymb}
\usepackage{setspace}
\usepackage[cp1251]{inputenc}

\begin{document}

\title{Tunable Flux through a Synthetic Hall Tube of Neutral Fermions}
\author{Xi-Wang Luo}
\affiliation{Department of Physics, The University of Texas at Dallas, Richardson, Texas
75080-3021, USA}
\author{Jing Zhang}
\thanks{Corresponding author. \\
Email: \href{mailto:jzhang74@sxu.edu.cn}{jzhang74@sxu.edu.cn}}
\affiliation{State Key Laboratory of Quantum Optics and Quantum Optics Devices, Institute
of Opto-Electronics, Shanxi University, Taiyuan 030006, P.R.China}
\affiliation{Synergetic Innovation Center of Quantum Information and Quantum Physics,
University of Science and Technology of China, Hefei, Anhui 230026, P. R.
China}
\author{Chuanwei Zhang}
\thanks{Corresponding author. \\
Email: \href{mailto:chuanwei.zhang@utdallas.edu}{chuanwei.zhang@utdallas.edu}%
}
\affiliation{Department of Physics, The University of Texas at Dallas, Richardson, Texas
75080-3021, USA}

\begin{abstract}
Hall tube with a tunable flux is an important geometry for studying quantum
Hall physics, but its experimental realization in real space is still
challenging. Here, we propose to realize a synthetic Hall tube with tunable
flux in a one-dimensional optical lattice with the synthetic ring dimension
defined by atomic hyperfine states. We investigate the effects of the flux
on the system topology and study its quench dynamics. Utilizing the tunable
flux, we show how to realize topological charge pumping, where interesting
charge flow and transport are observed in rotated spin basis. Finally, we
show that the recently observed quench dynamics in a synthetic Hall tube can
be explained by the random flux existing in the experiment.
\end{abstract}

\maketitle

\section{Introduction}

Ultracold atoms are emerging as a promising platform for the study of
condensed matter physics in a clean and controllable environment~\cite%
{Jaksch2005The,RevModPhys.80.885}. The capability of generating artificial
gauge fields and spin-orbit coupling using light-matter interaction~\cite%
{lin2011spin,zhang2012collective,qu2013observation,ji2014experimental,olson2014tunable,wang2012spin,cheuk2012spin,meng2016experimental, huang2016experimental,PhysRevA.98.013615,wu2016realization,campbell2015itinerant,luo2016tunable,kolkowitz2017spin,bromley2018dynamics,arXiv.1907.08637}
offers new opportunity for exploring topologically nontrivial states of
matter~\cite%
{dalibard2011colloquium,goldman2014light,Galitski2013Spin,Zhai2012Spin,Zhai2015Degenerate,Zhang2018Topological,zhang2014fermi}%
. One recent notable achievement was the realization of Harper-Hofstadter
Hamiltonian, an essential model for quantum Hall physics, using
laser-assisted tunneling for generating artificial magnetic fields in
two-dimensional (2D) optical lattices~\cite%
{PhysRevLett.108.225304,PhysRevLett.111.185301,PhysRevLett.111.185302}.
Moreover, synthetic lattice dimension defined by atomic internal states \cite%
{PhysRevLett.108.133001,PhysRevLett.112.043001,PhysRevLett.115.195303,arXiv.1809.02122,arXiv.1810.12331,mancini2015observation,Stuhl2015Visualizing,PhysRevLett.117.220401,PhysRevLett.122.065303,PhysRevResearch.2.013149}
provides a new powerful tool for engineering new high-dimensional quantum
states of matter with versatile boundary manipulation\ \cite%
{arXiv.1809.02122,arXiv.1810.12331}, using a low-dimensional physical system.

Nontrivial lattice geometries with periodic boundaries (such as a torus or
tube) allow the study of many interesting physics such as the Hofstadter's
butterfly~\cite{PhysRevB.14.2239} and Thouless pump~\cite%
{PhysRevB.23.5632,thouless1983quantization,PhysRevLett.64.1812,RevModPhys.82.1959,PhysRevLett.115.095302,PhysRevLett.118.230402}%
, where the flux through the torus or tube is crucially important. In a
recent experiment \cite{PhysRevLett.122.065303}, a synthetic Hall tube has
been realized in a 1D optical lattice and interesting quench dynamics have
been observed, where the flux effect was not considered. More importantly,
the flux through the tube, determined by the relative phase between Raman
lasers, is spatially non-uniform and random for different iterations of the
experiment, yielding major deviation from the theoretical prediction. The
physical significance and experimental progress raise two natural questions:
can the flux through the synthetic Hall tube be controlled and tuned in
realistic experiments? If so, can such controllability lead to the
observation of interesting quench dynamics?

In this paper, we address these important questions by proposing a simple
scheme to realize a controllable flux $\Phi $ through a three-leg synthetic
Hall tube and studying its quench dynamics and topological pumping. Our main
results are:

i) We use three hyperfine ground spin states, each of which is dressed by
one far-detuned Raman laser, to realize the synthetic ring dimension of the
tube. The flux $\Phi $ can be controlled simply by varying the polarizations
of the Raman lasers~\cite{meng2016experimental}. The scheme can be applied
to both Alkali (e.g., potassium)~\cite%
{meng2016experimental,PhysRevA.98.013615,huang2016experimental,kolkowitz2017spin,bromley2018dynamics}
and Alkaline-earth(-like) atoms (e.g., strontium, ytterbium)\ \cite%
{PhysRevA.76.022510}.

ii) The three-leg Hall tube is characterized as a 2D topological insulator
with $\Phi $ playing the role of the momentum along the synthetic dimension.
The system also belongs to a 1D topological insulator at $\Phi =0$ and $\pi $%
, where the winding number is quantized and protected by a generalized
inversion symmetry.

iii) The tunable $\Phi $ allows the experimental observation of topological
charge pumping in the tube geometry, which also probes the system topology.
Interesting charge flow and transport can be observed in rotated spin basis.

iv) We study the quench dynamics with a tunable flux and show that the
experimental observed quench dynamics in~\cite{PhysRevLett.122.065303} can
be better understood using a random flux existing in the experiment.


\section{The model}

\begin{figure}[t]
\includegraphics[width=1.0\linewidth]{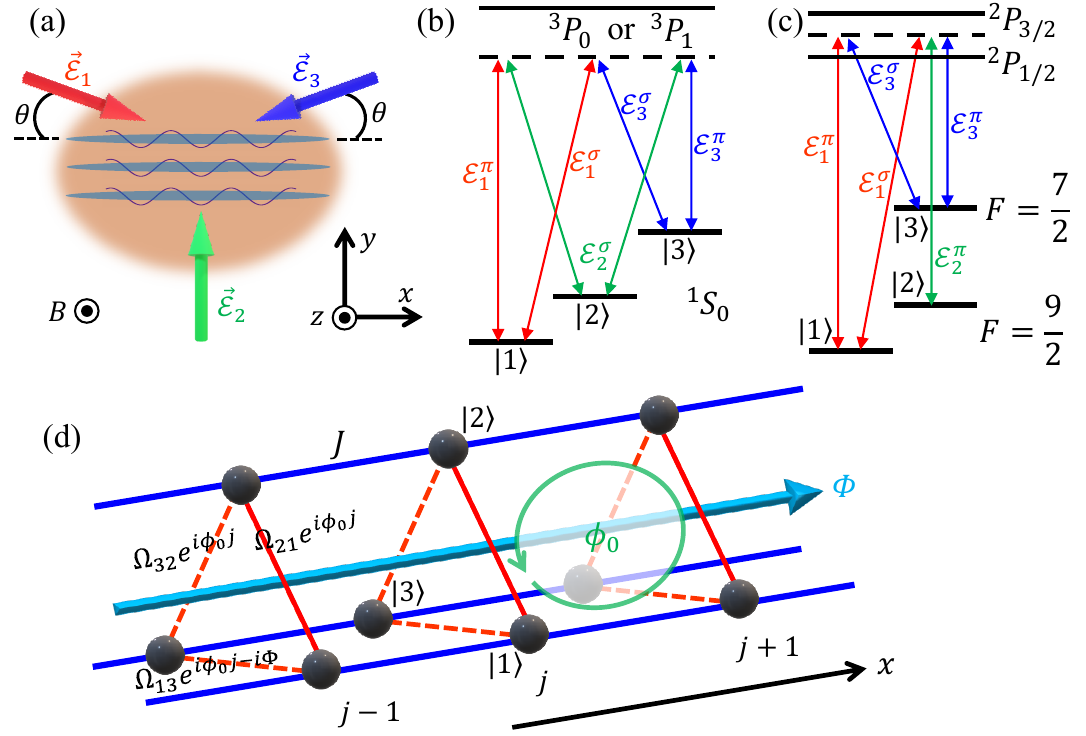}
\caption{(a) Schematic of the experimental setup for tunable flux through
the synthetic Hall tube. The Raman lasers $\vec{\mathcal{E}}_{1}$, $\vec{%
\mathcal{E}}_{2}$ and $\vec{\mathcal{E}}_{3}$ generate the couplings along
the synthetic dimension spanned by atomic hyperfine states. (b) and (c) The
corresponding Raman transitions for alkaline-earth(-like) atoms and alkali,
respectively. (d) Synthetic Hall tube with a uniform flux $\protect\phi _{0}$
on each side plaquette and $\Phi $ through the tube.}
\label{fig:sys1}
\end{figure}

We consider an experimental setup with cold atoms trapped in 1D optical
lattices along the $x$-direction, where transverse dynamics are suppressed
by deep optical lattices, as shown in Fig.~\ref{fig:sys1}a. The bias
magnetic field is along the $z$ direction to define the quantization axis.
Three far-detuned Raman laser fields $\vec{\mathcal{E}}_{s}$, propagating in
the $x$-$y$ plane, are used to couple three atomic hyperfine ground spin
states, with each state dressed by one Raman laser, as shown in Figs.~\ref%
{fig:sys1}b and \ref{fig:sys1}c for alkaline-earth(-like) (e.g., strontium,
ytterbium) and alkali (e.g., potassium) atoms, respectively. The three spin
states form three legs of the synthetic tube system as shown in Fig.~\ref%
{fig:sys1}d, and the tight-binding Hamiltonian is written as%
\begin{equation}
H=\sum_{j;s\neq s^{\prime }}\widetilde{\Omega }_{ss^{\prime
};j}c_{j,s}^{\dag }c_{j,s^{\prime }}-\sum_{j;s}(Jc_{j,s}^{\dag
}c_{j+1,s}+H.c.),  \label{Ham_real}
\end{equation}%
where $\widetilde{\Omega }_{ss^{\prime };j}=\Omega _{ss^{\prime }}e^{i\phi
_{j;ss^{\prime }}}$, $c_{j,s}^{\dag }$ is the creation operator with $j$, $s$
the site and spin index. $J$ and $\Omega _{ss^{\prime }}$ are the tunneling
rate and Raman coupling strength, respectively. For alkaline-earth(-like)
atoms, we use three states in the $^{1}S_{0}$ ground manifold to define the
synthetic dimension. The long lifetime $^{3}P_{0}$ or $^{3}P_{1}$ levels are
used as the intermediate states for the Raman process (see Fig.~\ref%
{fig:sys1}b) such that $\delta m_{F}=\pm 2$ Raman process does not suffer
the heating issues~\cite{PhysRevLett.117.220401}. While for alkaline atoms
(see Fig.~\ref{fig:sys1}c), we choose three hyperfine spin states $%
|F,m_{F}\rangle $, $|F,m_{F}-1\rangle $ and $|F-1,m_{F}\rangle $ from the
ground-state manifold to avoid $\delta m_{F}=\pm 2$ Raman process, so that
far-detuned Raman lasers tuned between the $D1$ and $D2$ lines can be used
to reduce the heating~\cite{meng2016experimental,huang2016experimental}.

The laser configuration in Fig.~\ref{fig:sys1}a generates a uniform magnetic
flux penetrating each side plaquette as well as a tunable flux through the
tube. Each Raman laser may contain both $z$-polarization (responsible for $%
\pi $ transition) and in-plane-polarization (responsible for $\sigma $
transition) components, which can be written as $\vec{\mathcal{E}}_{s}=\hat{%
\mathbf{e}}_{\pi }\mathcal{E}_{s}^{\pi }+\hat{\mathbf{e}}_{\sigma }\mathcal{E%
}_{s}^{\sigma }$. For alkaline-earth(-like) atoms, we choose $\mathcal{E}%
_{2}^{\pi }=0$ so that $\widetilde{\Omega }_{21;j}\propto \mathcal{E}%
_{1}^{\pi }\mathcal{E}_{2}^{\sigma \ast}$, $\widetilde{\Omega }%
_{32;j}\propto \mathcal{E}_{2}^{\sigma }\mathcal{E}_{3}^{\pi \ast}$ and $%
\widetilde{\Omega }_{13;j}\propto \mathcal{E}_{3}^{\sigma }\mathcal{E}%
_{1}^{\sigma \ast}$. Therefore, the corresponding Raman coupling phases are $%
\phi _{j;21}=(\mathbf{k}_1-\mathbf{k}_2)\cdot\mathbf{x}_j+\varphi _{1}^{\pi
}-\varphi _{2}^{\sigma }$, where $\mathbf{k}_s$ is the wave vector of the $s$%
-th Raman laser, $\varphi _{s}^{\pi ,\sigma }$ are the $(\pi ,\sigma )$%
-component phases of the $s$-th Raman laser at site $j=0$ and $\mathbf{x}%
_j=jd_x \hat{\mathbf{x}}$ is the position of site $j$ with $d_{x}$ the
lattice constant. As a result, we have $\phi _{j;21}=j\phi _{0}+\varphi
_{1}^{\pi }-\varphi _{2}^{\sigma }$ where $\phi _{0}=(\mathbf{k}_1-\mathbf{k}%
_2)\cdot\hat{\mathbf{x}} =k_{R}d_{x}\cos (\theta ) $ gives rise to the
magnetic flux penetrating the side plaquette of the tube, with $k_{R}$ the
recoil momentum of the Raman lasers. Similarly, we have $\phi _{j;32}=j\phi
_{0}+\varphi _{2}^{\sigma }-\varphi _{3}^{\pi }$, $\phi _{j;13}=-2j\phi
_{0}+\varphi _{3}^{\sigma }-\varphi _{1}^{\sigma }$. To obtain a
uniform magnetic flux for each side plaquette of the tube, we choose the
incident angle $\theta $ such that the magnetic flux for each side plaquette
is $\phi _{0}=2\pi /3$. The phases $\varphi _{s}^{\sigma ,\pi }$
determine the flux through the tube which is $\Phi\equiv-\phi_{j;21}-\phi
_{j;32}-\phi _{j;13}=\varphi _{3}^{\sigma }-\varphi _{3}^{\pi }+\varphi
_{3}^{\pi }-\varphi _{3}^{\sigma }$. Therefore, we obtain $\Phi =\Delta
\varphi _{3}-\Delta \varphi _{1}$, where $\Delta \varphi _{s}=\varphi
_{s}^{\pi }-\varphi _{s}^{\sigma }$ is the phase difference between two
polarization components of the $s$-th Raman laser. We notice that the phase
of each Raman coupling (i.e., $\phi_{j;ss^{\prime }}$) depends on random
phase difference between two Raman lasers (e.g., $\phi_{j;21}$ depends on $%
\varphi _{1}^{\pi }-\varphi _{2}^{\sigma}$); however, their summation (i.e.,
the flux $\Phi$) only depends on the phase difference $\Delta \varphi _{s}$
between two polarizations of the same Raman laser, which can be controlled
at will using wave plates. Moreover, the phase differences $\Delta \varphi
_{s}$ do not depend on the transverse positions $y$ and $z$. Basically, the
reason why we can control the flux $\Phi$ is that each spin state is dressed
by one and only one Raman laser, and the random global phases of the Raman
lasers can be gauged out by absorbing it into the definition of the spin
states. With proper gauge choice, we can set the tunneling phases as $\phi
_{j;21}=\phi _{j;32}=j\phi _{0}$ and $\phi _{j;13}=j\phi _{0}+\Phi $, as
shown in Fig.~\ref{fig:sys1}d.

For alkali atoms, we choose $\mathcal{E}_{2}^{\sigma }=0$, yielding $%
\widetilde{\Omega }_{21;j}=\alpha _{21}\mathcal{E}_{1}^{\sigma }\mathcal{E}%
_{2}^{\pi \ast}$, $\widetilde{\Omega }_{32;j}=\alpha _{32}\mathcal{E}%
_{2}^{\pi }\mathcal{E}_{3}^{\pi \ast}$ and $\widetilde{\Omega }_{13;j}=\beta
_{13}\mathcal{E}_{3}^{\pi }\mathcal{E}_{1}^{\sigma \ast}+\alpha _{13}%
\mathcal{E}_{3}^{\sigma }\mathcal{E}_{1}^{\pi \ast}$, with $\alpha
_{s,s^{\prime }},\beta _{s,s^{\prime }}$ determined by the transition dipole
matrix. We further consider $(\mathcal{E}_{3}^{\pi },\mathcal{E}_{1}^{\sigma
})\ll (\mathcal{E}_{3}^{\sigma },\mathcal{E}_{1}^{\pi })\ll \mathcal{E}%
_{2}^{\pi }$ and $\mathcal{E}_{3}^{\sigma }\mathcal{E}_{1}^{\pi }\simeq
\mathcal{E}_{3}^{\pi }\mathcal{E}_{2}^{\pi }\simeq \mathcal{E}_{1}^{\sigma }
\mathcal{E}_{2}^{\pi }$. Therefore, we have $\widetilde{\Omega }%
_{13;j}\simeq \alpha _{13}\mathcal{E}_{1}^{\pi \ast}\mathcal{E}_{3}^{\sigma }
$ with amplitudes $\Omega _{21}\simeq \Omega _{32}\simeq \Omega _{13}$.
Similar as the alkaline-earth(-like) atoms, we obtain uniform magnetic flux $%
\phi _{0}=2\pi /3$ for tube side by choosing $k_{R}d_{x}\cos (\theta )=2\pi
/3$. The flux through the tube becomes $\Phi =\Delta \varphi _{3}+\Delta
\varphi _{1}$, which can also be tuned at will through the polarization
control. 


\section{Phase diagram}

\begin{figure}[t]
\includegraphics[width=1.0\linewidth]{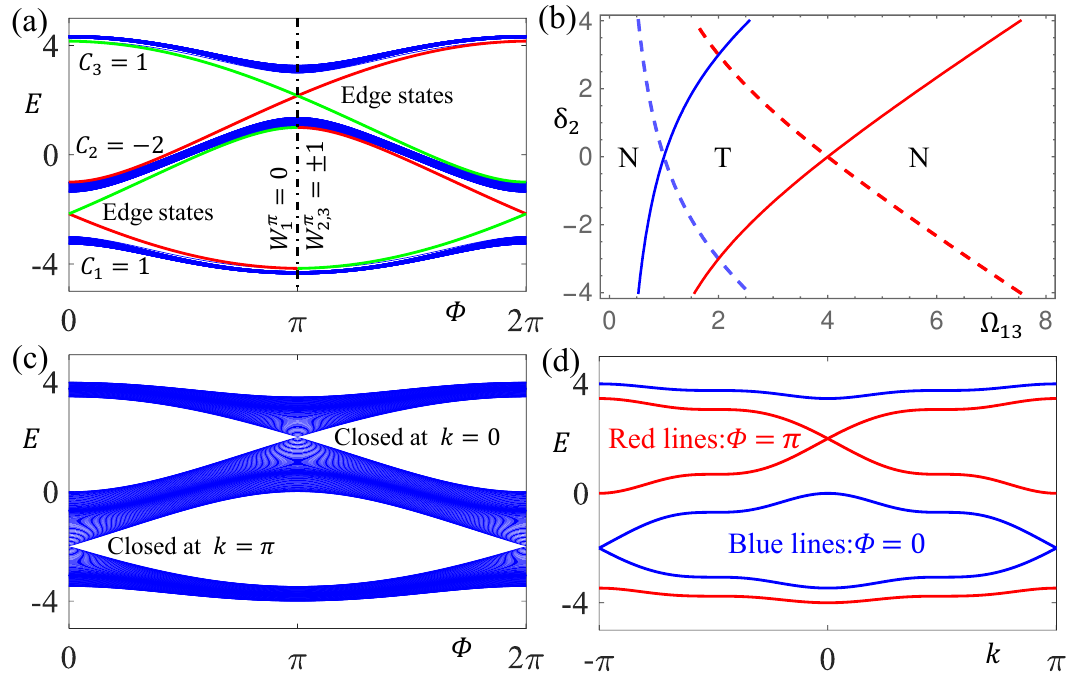}
\caption{(a) Band structures in the topological phase. (b) Phase diagram in
the $\Omega _{13}$-$\protect\delta _{2}$ plane, with solid (dashed) lines
the boundary between topological (T) and normal (N) phases for the upper
(lower) gap. (c) and (d) Band structures at the left phase boundary with $%
\protect\delta _{2}=0$ and $\Omega _{13}=1$. Common parameters: $\Omega =2$
with energy unit $J$.}
\label{fig:phase}
\end{figure}

The Bloch Hamiltonian in the basis $[c_{k,1},c_{k,2},c_{k,3}]^{T}$ reads
\begin{equation}
H_{k}=\left[
\begin{array}{ccc}
-2J\cos (k-\phi _{0}) & \Omega _{21} & \Omega _{13}e^{-i\Phi } \\
\Omega _{21} & -2J\cos (k) & \Omega _{32} \\
\Omega _{13}e^{i\Phi } & \Omega _{32} & -2J\cos (k+\phi _{0})%
\end{array}%
\right] ,  \label{Ham_k}
\end{equation}%
with $k$ the momentum along the real-space lattice. For $\Omega _{21}=\Omega
_{32}=\Omega _{13}$, the above Hamiltonian is nothing but the
Harper-Hofstadter Hamiltonian with $\Phi $ the effective momentum along the
synthetic dimension, and $\phi _{0}=2\pi /3$ is the flux per plaquette. The
topology is characterized by the Chern number~\cite{PhysRevLett.49.405}
\begin{equation}
C_{n}=\frac{i}{2\pi }\int dkd\Phi \langle \partial _{\Phi }u_{n}|\partial
_{k}|u_{n}\rangle -\langle \partial _{k}u_{n}|\partial _{\Phi }|u_{n}\rangle
,
\end{equation}%
where $|u_{n}\rangle $ is the Bloch states of the $n$-th band, satisfying $%
H_{k}({\Phi })|u_{n}(\Phi ,k)\rangle =E_{n}(\Phi ,k)|u_{n}(\Phi ,k)\rangle $%
. In Fig.~\ref{fig:phase}a, we plot the band structures as a function of $%
\Phi $ with an open boundary condition along the real-lattice direction.
There are three bands, and two gapless edge states (one at each ends) in
each gap. The two edge states cross only at $\Phi =0$ and $\Phi =\pi $,
where the tube belongs to a 1D topological insulator with quantized winding
number (Zak phase)~\cite{PhysRevLett.62.2747}
\begin{equation}
W_{n}^{0,\pi }=\frac{1}{\pi }\oint dk\langle u_{n}|\partial
_{k}|u_{n}\rangle \big|_{\Phi =0,\pi }.
\end{equation}%
The winding number is protected by a generalized inversion symmetry $%
\mathcal{I}H_{k}\mathcal{I}^{-1}=H_{-k}$, where the inversion symmetry $%
\mathcal{I}$ swaps spin states $|1\rangle $ and $|3\rangle $~\cite%
{PhysRevA.97.013634}.

The Chern number and winding number are still well defined even when the
Raman couplings have detunings and/or the coupling strength $\Omega
_{ss^{\prime }}$ are nonequal. The changes in these Raman coupling
parameters drive the phase transition from topological to trivial
insulators. The detuning can be introduced by including additional terms $%
\sum_{j;s}\delta _{s}c_{j,s}^{\dag }c_{j,s}$ in the Hamiltonian Eq.~\ref%
{Ham_real}. Hereafter we will fix $\Omega _{21}=\Omega _{32}\equiv \Omega $
and $\delta _{1}=\delta _{3}=0$ for simplicity. The phase diagram in the $%
\Omega _{13}$-$\delta _{2}$ plane is shown in Fig.~\ref{fig:phase}b. The
solid (dashed) lines are the phase boundaries corresponding to the gap
closing between two lower (upper) bands, with topological phases between two
boundaries. At the phase boundaries, the corresponding band gaps close at $%
\Phi =0$ ($\Phi =\pi $) for the two lower (upper) bands, as shown in Fig.~%
\ref{fig:phase}c. In addition, for the two lower (upper) bands, the gap
closes at $k=0$ and $k=\pi $ ($k=\pi $ and $k=0$) on the right and left
boundaries, respectively, as shown in Fig.~\ref{fig:phase}d. The gaps reopen
in the trivial phase with the disappearance of edge states.

\begin{figure}[t]
\includegraphics[width=1.0\linewidth]{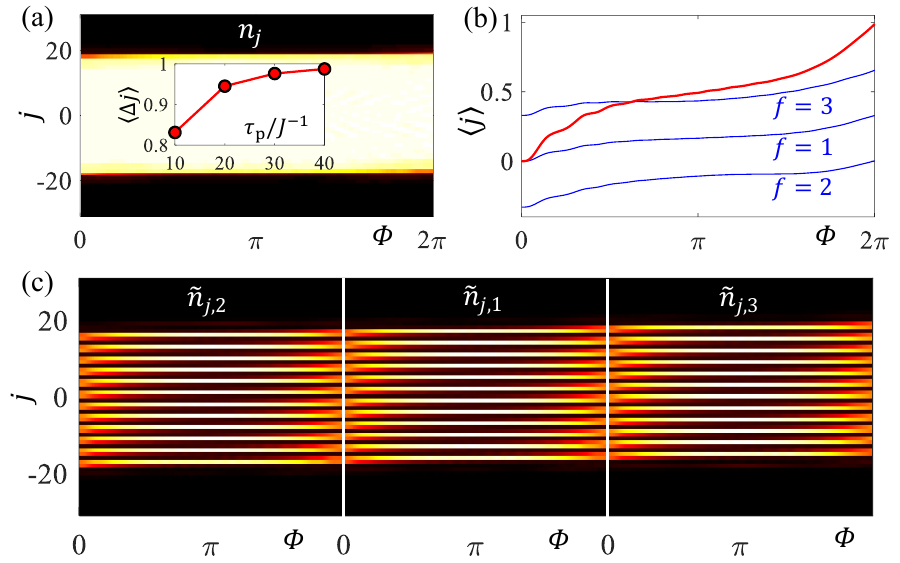}
\caption{(a) Total density distribution during one pump cycle. Inset shows
the non-adiabatic effects on the center-of-mass shift. (b) Center-of-mass
(red line) and the rotated-spin contributions (blue lines) during one pump
cycle. The blue lines are $\sum_{j}j\widetilde{n}_{j,f}(t)/N$ with $f=1,2,3$
as labeled. (c) Rotated spin-density distributions during one pump cycle.}
\label{fig:pump}
\end{figure}

\section{Topological pumping}

The three-leg Hall tube is a minimal Laughlin's cylinder. When the flux
through the tube is adiabatically changed by $2\pi $, the shift of
Wannier-function center is proportional to the Chern number of the
corresponding band~\cite%
{PhysRevB.23.5632,thouless1983quantization,PhysRevLett.64.1812,RevModPhys.82.1959}%
. Therefore all particles are pumped by $C$ site (with $C$ the total Chern
number of the occupied bands) as $\Phi $ changes by $2\pi $, i.e., $C$
particles are pumped from one edge to another. Given the ability of
controlling the flux through the tube, we can measure the topological Chern
number based on topological pumping by tuning the flux $\Phi $ adiabatically
(compared to the band gaps).

Here we consider the Fermi energy in the first gap with only the lowest $C=1$
band occupied, and study the zero temperature pumping process (the pumped
particle is still well quantized for low temperature comparing to the band
gap)~\cite{PhysRevLett.111.026802}. The topological pumping effect can be
identified as the quantized center-of-mass shift of the atom cloud~\cite%
{PhysRevLett.110.166802,PhysRevLett.111.026802,lohse2015thouless,nakajima2016topological}
in a weak harmonic trap $V_{\text{trap}}=\frac{1}{2}v_{\text{T}}\ j^{2}$.
The harmonic trap strength $v_{\text{T}}=0.008J$ and the atom number $N=36$
are chosen such that the atom cloud has a large insulating region with one
atom per unit-cell at the trap center (In a realistic experiment, the above
parameters are typical). For simplicity, we set $\Omega _{ss^{\prime }}=J$
and $\delta _{s}=0$ for all $s,s^{\prime }$, choose the gauge as $\phi
_{j;21}=\phi _{j;32}=j\phi _{0}$, $\phi _{j;13}=j\phi _{0}-\Phi $, and
change $\Phi $ slowly (compared to the band gap) as $\Phi (t)=\frac{2\pi t}{%
\tau _{\text{p}}}$.
In Figs.~\ref{fig:pump}a and \ref{fig:pump}b, we plot the total density
distribution $n_{j}(t)$ and the center-of-mass $\langle j(t)\rangle \equiv
\sum_{j}jn_{j}(t)/N$ shift during one pumping circle with $\tau _{\text{p}%
}=40J^{-1}$, and we clearly see the quantization of the pumped atom $\langle
\Delta j\rangle \equiv \langle j(\tau _{\text{p}})\rangle -\langle
j(0)\rangle =1$. The atom cloud shifts as a whole with $n_{j}(t)=1$ near the
trap center. The inset in Fig.~\ref{fig:pump}a shows non-adiabatic effect
(finite pumping duration $\tau _{\text{p}}$) on the pumped atom. Typically, $%
J/2\pi$ is about several hundred Hz; therefore, the pumping duration $\tau _{%
\text{p}}$ should be of the order of tens of ms to satisfy the adiabatic
condition. This timescale is of the same magnitude as the lattice loading
and time-of-flight imaging duration~\cite%
{mancini2015observation,Stuhl2015Visualizing}. Starting from a degenerate
Fermi gas prepared at the optical dipole trap, the whole experimental
duration is less than 1s (which is typical for cold atom experiments).

The atoms are equally distributed on the three spin states [i.e., $%
n_{j,s}(t)\equiv \langle c_{j,s}^{\dag }c_{j,s}\rangle =\frac{1}{3}n_{j}(t)$%
]. To see the pumping process more clearly, we can examine the spin
densities in the rotated basis $\widetilde{n}_{j,f}(t)\equiv \langle
\widetilde{c}_{j,f}^{\dag }\widetilde{c}_{j,f}\rangle $, with $\widetilde{c}%
_{j,f}=\frac{1}{\sqrt{3}}\sum_{s}c_{j,s}e^{is\frac{2f\pi -\Phi }{3}}$. The
Hamiltonian in these basis reads $H=\sum_{f=1}^{3}H_{f}$, where
\begin{equation}
H_{f}=2J\cos (K_{j,f,\Phi })\widetilde{c}_{j,f}^{\dag }\widetilde{c}_{j,f}+(J%
\widetilde{c}_{j,f}^{\dag }\widetilde{c}_{j,f+1}+H.c.)
\end{equation}%
is the typical Aubry-Andr{\'{e}}-Harper (AAH) Hamiltonian~\cite{AAH1,AAH2}
with $K_{j,f,\Phi }=\frac{2\pi (f+j)-\Phi }{3}$. Each spin component
contributes exactly one third of the quantized center-of-mass shift (see
Fig.~\ref{fig:pump}b). The bulk-atom flow during the pumping can be clearly
seen from the spin densities $\widetilde{n}_{j,f}(t)$, as shown in Fig.~\ref%
{fig:pump}c. The quantized pumping can also be understood by noticing that
the AAH Hamiltonians $H_{s}$ are permutated as $H_{1}\rightarrow
H_{3}\rightarrow H_{2}\rightarrow H_{1}$ after one pump circle. Each $H_{f}$
returns to itself after three pump circles with particles pumped by three
sites (since the lattice period of $H_{f}$ is 3). Therefore, for the total
Hamiltonian $H$, particles are pumped by one site after one pump circle. The
physics for different values of $\Omega _{ss^{\prime }}$ and $\delta _{s}$
are similar, except that the rotated basis $\widetilde{c}_{j,f}$ may take
different forms.

In the presence of atom-atom interactions, the robust topological properties
should not be affected if the interaction strength is much weaker compared
to the band gaps, and the charge pumping should remain quantized~\cite%
{Niu1984Quantised}. The interaction can be written as $H_\text{int}=\frac{U}{%
2}\sum_{j,s\neq s^{\prime }}\hat{n}_{j,s}\hat{n}_{j,s^{\prime }}$ with atom
number operator $\hat{n}_{j,s}=c^\dag_{j,s}c_{j,s}$ and interaction strength
$U$. Atom in site $(j,s)$ would interact with atoms in all other sites $%
(j,s^{\prime }\neq s)$ along the synthetic dimension. Therefore, the
interaction is long ranged along the synthetic dimension.
At the strong interaction region, the system may produce fractional quantum
Hall physics and support fractional topological pumping~\cite%
{PhysRevLett.115.095302,PhysRevLett.118.230402}. Our scheme generates a
tunable flux through the tube and thus offers an ideal platform for studying
quantized fractional charge transport and probing the fractional many-body
Chern numbers.


\begin{figure}[t]
\includegraphics[width=1.0\linewidth]{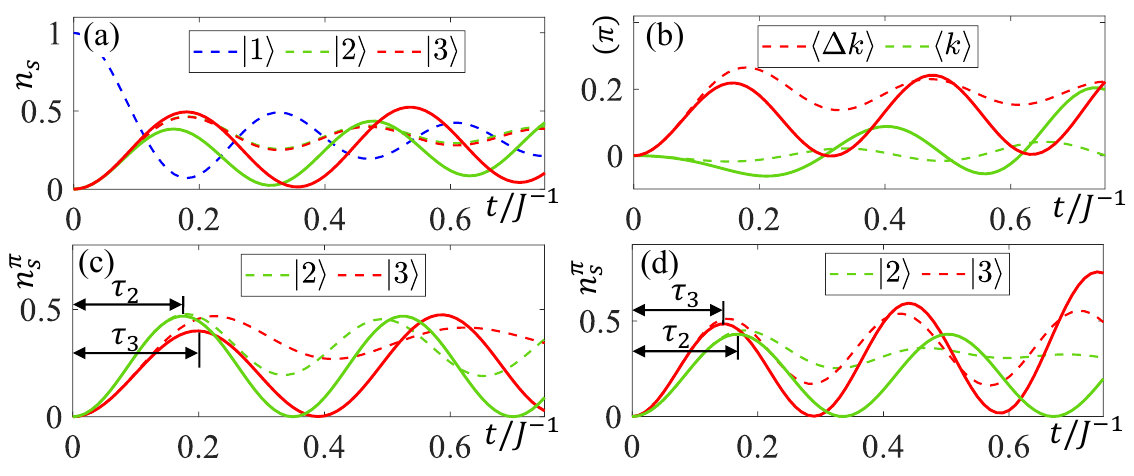}
\caption{Quench dynamics for $\Phi =0$ (solid lines) and averaged over
random $\Phi $ (dashed lines). Time evolution of spin populations (a) and
averaged momenta (b) with $\Omega _{13}=\Omega $. Time evolution of spin
populations at $k=\protect\pi $ with $\Omega _{13}=5$ in (c) and $\Omega
_{13}=7.5$ in (d). $n_{s}^{\protect\pi }=\frac{n_{s}(\protect\pi )}{%
\sum_{s^{\prime }}n_{s^{\prime }}(\protect\pi )}$, $\protect\tau _{2}$ and $%
\protect\tau _{3}$ cross at $\Omega _{13}^{c}=5.8$. Both gaps are
topological in (a) and (b), and only the upper (lower) gap is topological in
(c) [(d)]. Common parameters: $\Omega =6.15$, $\protect\delta %
_{2}=-0.4\Omega $ with energy unit $J$.}
\label{fig:quench}
\end{figure}

\section{Quench dynamics}

Besides topological pumping, the quench dynamics of the system can also be
used to demonstrate the presence of gauge field $\phi _{0}$ and detect the
phase transitions~\cite{PhysRevLett.122.065303}. Here we study how $\Phi $
affects the quench dynamics by
considering that all atoms are initially prepared in state $|1\rangle $,
then the inter leg couplings are suddenly activated by turning on the Raman
laser beams.
In Figs.~\ref{fig:quench}a and \ref{fig:quench}b, we show the time evolution
of the fractional spin populations $n_{s}=\frac{1}{N}\int dkn_{s}(k)$, as
well as the momenta $\langle k\rangle =\sum_{s}\langle k_{s}\rangle $ and $%
\langle \Delta k\rangle =\langle k_{2}\rangle -\langle k_{3}\rangle $ (both
can be measured by time-of-flight imaging) for $\Phi =0$,
where $n_{s}(k)=\langle c_{s}(k)^{\dag }c_{s}(k)\rangle $ and $\langle
k_{s}\rangle =\frac{1}{N}\int kn_{s}(k)dk$ with $N$ the total atom number.
We find that the time evolutions show similar oscillating behaviors for
different $\Phi $, but with different frequencies and amplitudes. The
difference between the momenta of atoms transferred to state $|2\rangle $
and $|3\rangle $ increase noticeably at early time as a result of the
magnetic flux $\phi _{0}$ penetrating the surface of the tube~\cite%
{PhysRevLett.122.065303}, which does not depend on the flux $\Phi $ through
the tube. In Fig.~\ref{fig:quench}, we have fixed $\Omega =6.15J$, $\delta
_{2}=-0.4\Omega $ and $U=0$. The initial temperature is set to be $%
T/T_{F}=0.3$, with initial Fermi temperature $T_{F}$ given by the difference
between the Fermi energy $E_{F}$ and the initial band minimum $-2J$ (i.e., $%
T_{F}=E_{F}+2J$). We use $T_{F}=2J$ to get the similar initial filling and
Fermi distribution as those in the experiment~\cite{PhysRevLett.122.065303}
(the results are insensitive to $T_{F}$).

\begin{figure}[tb]
\includegraphics[width=1.0\linewidth]{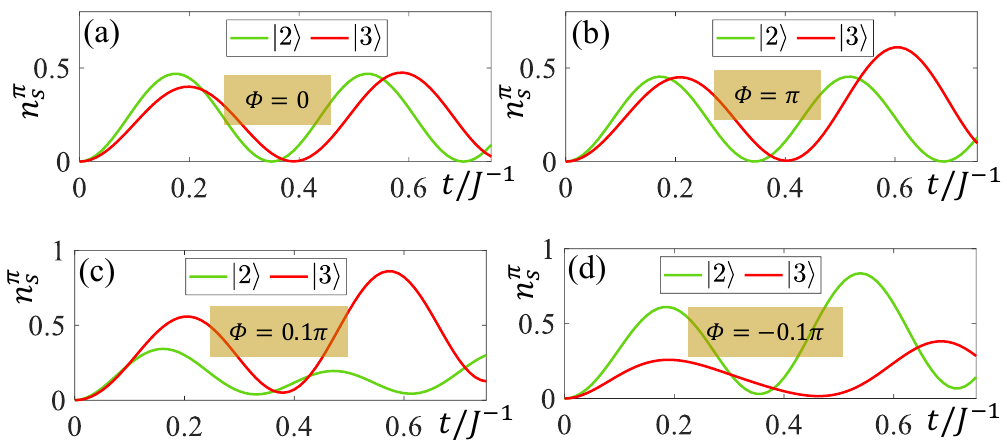}
\caption{(a)-(d) Time evolution of the spin populations at $k=\protect\pi$
for different values of $\Phi$. All other parameters are the same as Fig.~%
\protect\ref{fig:quench}c.}
\label{fig:APPPhi}
\end{figure}

The quench dynamics can also be used to measure the gap closing at phase
boundaries.
Similar as Ref.~\cite{PhysRevLett.122.065303}, we introduce two times $\tau
_{2}$ and $\tau _{3}$, at which the spin-$|2\rangle $ and spin-$|3\rangle $
populations at $k=0$ (or $k=\pi $) reach their first maxima, to identify the
phase boundary. As we change $\delta _{2}$ or $\Omega _{13}$ across the
phase boundary (one gap closes and the dynamics is characterized by a single
frequency), $\tau _{2}$ and $\tau _{3}$ cross each other. Notice that above
discussions only apply to $\Phi =0,\pi $ where the gap closing occurs. We
find that, even when no gap closing occurs for $\Phi$ away from $0$ and $\pi$%
, $\tau_2$ and $\tau_3$ would cross each other as we increase $\Omega_{13}$,
and the crossing point is generally away from the phase boundaries.
Therefore, the measurement of gap closing based on quench dynamics is
possible only if $\Phi $ can be controlled. As an example, we consider $\Phi
=0$ and plot the time evolution of the spin populations at $k=\pi $ with $%
\Omega _{13}$ around the left phase boundary $\Omega _{13}^{c}$, as shown in
Figs.~\ref{fig:quench}c and \ref{fig:quench}d. The crossing points are
different for different $\Phi$. In Fig.~\ref{fig:APPPhi}, we plot the time
evolution of the spin populations at $k=\pi$ for different values of $\Phi$.

The ability to control the flux $\Phi $ is crucial for the study of both
topological properties and quench dynamics. We notice that for the
experiment in~\cite{PhysRevLett.122.065303}, $\Phi $ cannot be controlled
and may vary from one experimental realization to another, $\Phi $ is also
different for different tubes within one experimental realization.
It is straightforward to show that the flux $\Phi$ in~\cite%
{PhysRevLett.122.065303} is determined by the difference between the random
phases of the Raman lasers at lattice sites (see Appendix),
which cannot be controlled since the Raman lasers propagate along different
paths, not to mention that their wavelengths are generally not commensurate
with the lattice. Moreover, in realistic experiments, arrays of independent
fermionic synthetic tubes are realized simultaneously due to the transverse
atomic distributions in the $y$, $z$ directions~\cite%
{PhysRevLett.117.220401,PhysRevLett.122.065303}, and the synthetic tubes at
different $y$ would have different $\Phi $ due to the $y$-dependent $\varphi
_{1}^{\sigma }$. For the parameters in Ref.~\cite{PhysRevLett.122.065303},
the difference of $\Phi $ between neighbor tubes in $y$ direction is about $%
2\pi \times 0.58$. The random flux or phase problem is a common issue for
schemes in which one spin state is dressed by two or more different lasers
(different in wave vector) where the random global phases of the Raman
lasers can not be gauged out.

Due to the randomness of $\Phi $, the dynamics in experiment~\cite%
{PhysRevLett.122.065303} (averaged over enough samplings) should correspond
to results averaged over $\Phi $. In Fig.~\ref{fig:quench}, we plot the
corresponding dynamics averaged over $\Phi $, which show damped oscillating
behaviors (with significant long-time damping), as observed in the
experiment. $\tau _{2}$ and $\tau _{3}$, which cross each other at different
values of $\Omega _{13}$ for different $\Phi $ with averaged value $\bar{%
\Omega}_{13}=\Omega \neq \Omega _{13}^{c}$, may not be suitable to identify
the phase boundaries (i.e., gap closings) for a random flux $\Phi$.

\begin{figure}[t]
\includegraphics[width=1.0\linewidth]{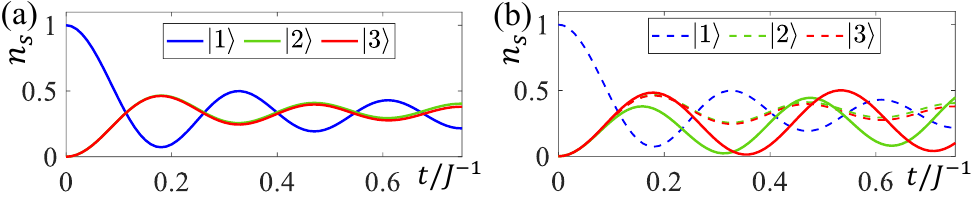}
\caption{Quench dynamics in the presence of a harmonic trap $V_\text{trap}=%
\frac{1}{2}v_\text{T} j^2$ for non-interacting fermions ($U=0$) in (a) and
interacting fermions ($U=1.7J$) in (b). The trap strength is $v_\text{T}%
\simeq0.0158J$. The results in (a) are averaged over $\Phi$. The solid
(dashed) lines in (b) are the results for $\Phi=0$ (averaged over $\Phi$).
All other parameters are the same as in Fig.~\protect\ref{fig:quench}a.}
\label{fig:AppTrap}
\end{figure}

In realistic experiments, a harmonic trap is usually applied to confine the
atoms. The above results for the quench dynamics still hold for atoms in a
weak harmonic trap, as confirmed by our numerical simulations. We consider a
harmonic trap frequency $\omega _{x}=2\pi \times 57$ Hz and $J=2\pi \times
264$ Hz as in the experiment~\cite{PhysRevLett.122.065303}, therefore, the
harmonic trap $V_{\text{trap}}=\frac{1}{2}v_{\text{T}}\ j^{2}$ has trap
strength $v_{\text{T}}\simeq 0.0158J$. In Fig.~\ref{fig:AppTrap}a, we plot
the time evolution of spin populations, we see that the quench dynamics are
hardly affected by the harmonic trap. Moreover, atom-atom interactions may
induce scattering between different momentum states that oscillate at
different frequencies, leading to additional damping, which should be minor
due to the short evolution time $\lesssim1$ms here. In Fig.~\ref{fig:AppTrap}%
b, we plot the time evolution of spin populations in the presence of weak
interaction $U=1.7J$~\cite{PhysRevLett.122.065303}. We have adopted the
mean-field approach and written the interaction as $\frac{U}{2}\sum_{j,s\neq
s^{\prime }}\hat{n}_{j,s}\hat{n}_{j,s^{\prime }}=\frac{U}{2}\sum_{j,s\neq
s^{\prime }}[\langle\hat{n}_{j,s^{\prime }}\rangle\hat{n}_{j,s}+\langle\hat{n%
}_{j,s}\rangle\hat{n}_{j,s^{\prime }} -\langle\hat{n}_{j,s}\rangle\langle%
\hat{n}_{j,s^{\prime }}\rangle]$, where the $s$-spin fermions interact with
the average density of the $s^{\prime }$-spin fermions, which should be
valid for weak interaction and short evolution time (similar to the Hartree
approximation). The weak interaction can hardly affect the quench dynamics
where the oscillation frequency ($\sim\Omega$) is much larger than the
interaction strength. Other experimental imperfections such as
spin-selective imaging error may also affect the measured spin dynamics, and
the final observed cross point in~\cite{PhysRevLett.122.065303} is smaller
than $\Omega $ and $\Omega _{13}^{c}$. %

\section{Conclusion}

In summary, we propose a simple and feasible scheme to realize a
controllable flux $\Phi $ through the synthetic Hall tube that can be tuned
at will, and study the effects of the flux $\Phi $ on the system topology
and dynamics. Previous experimental quench dynamics may be better explained
by the results averaged over the random flux existing in the experiment.
Our scheme allows the study of interesting charge transport (e.g., the
interesting charge flow and transport for rotated spin states in our system)
through topological pumping, which also probes the system topology.
Moreover, it may open the possibility to study fractional charge pumping and
probe the many-body Chern numbers of fractional quantum Hall state at strong
interactions. Our results provide a new platform for studying topological
physics in a tube geometry with tunable flux and may be generalized with
other synthetic degrees of freedom, such as momentum states~\cite%
{Gadway2016NC,Gadway2017SA,Gadway2017NC} and lattice orbitals~\cite%
{PhysRevA.95.023607,PhysRevLett.121.150403}.

\begin{acknowledgments}
\textbf{Acknowledgements}: XWL and CZ are supported by AFOSR
(FA9550-16-1-0387, FA9550-20-1-0220), NSF (PHY-1806227), and ARO
(W911NF-17-1-0128). JZ is supported by the National Key Research and
Development Program of China (2016YFA0301602).
\end{acknowledgments}

\section{Appendix: Random flux $\Phi$ in previous experiments}

\begin{figure}[h]
\includegraphics[width=1.0\linewidth]{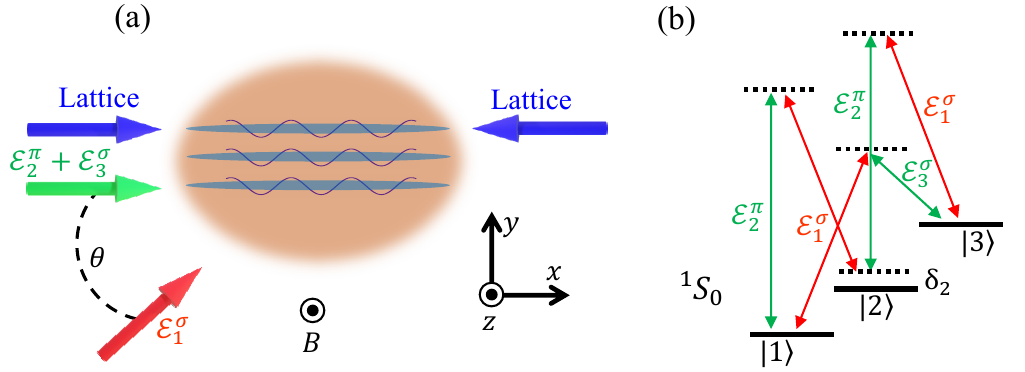}
\caption{(a) Schematic of the experimental setup and (b) The corresponding
Raman transitions in~\protect\cite{PhysRevLett.122.065303}.}
\label{fig:sys2S}
\end{figure}

We would like to emphasize that our proposed setup to generate a tunable
flux through the synthetic tube is different from the setup in recent
experimental work~\cite{PhysRevLett.122.065303} (as shown in Fig.~\ref%
{fig:sys2S} for comparison). The Raman laser configurations (directions,
polarizations and frequencies) as well as the involved Raman processes are
different. It is straightforward to show that the flux $\Phi$ for the setup
in Fig.~\ref{fig:sys2S} is $\Phi =3\varphi _{1}^{\sigma }-2\varphi _{2}^{\pi
}-\varphi _{3}^{\sigma }$, which cannot be controlled since the Raman lasers
propagate along different paths, not to mention that their wavelengths are
generally not commensurate with the lattice. The random flux or phase
problem is a common issue for schemes in which one spin state is dressed by
two or more different lasers (different in wave vector) where the random
global phases of the Raman lasers can not be gauged out. Moreover, in
realistic experiments, arrays of independent fermionic synthetic tubes are
realized simultaneously due to the transverse atomic distributions in the $y$%
, $z$ directions, and the synthetic tubes at different $y$ would have
different $\Phi $ due to the $y$-dependent $\varphi _{1}^{\sigma }$ for the
setup in Fig.~\ref{fig:sys2S}.

\end{document}